\documentclass[sigconf,nonacm]{acmart}

\acmConference[Responsible AI @ KDD]{KDD Workshop on Responsible AI}{2021}{Singapore}

\usepackage{color}

\usepackage{wrapfig}

\begin{document}

\title{Measurement as governance in and for responsible AI}

\author{Abigail Z.\ Jacobs}
\email{azjacobs@umich.edu}
\affiliation{
   \institution{University of Michigan}
}

\renewcommand{\shortauthors}{A.\ Z.\ Jacobs}

\begin{abstract}

Measurement of social phenomena is everywhere, unavoidably, in sociotechnical systems. 
This is not (only) an academic point:
Fairness-related harms emerge when there is a mismatch in the measurement process between the thing we purport to be measuring and the thing we actually measure.
However, the measurement process---where social, cultural, and political values are implicitly encoded in sociotechnical systems---is almost always obscured. \looseness=-1
Furthermore, this obscured 
process is where important governance decisions are encoded: governance about which systems are fair, which individuals belong in which categories, and so on.
We can then use the language of measurement, and the tools of construct validity and reliability, to uncover hidden governance decisions. 
In particular, we highlight two types of construct validity, content validity and consequential validity, that are useful to 
elicit and characterize the feedback loops between the measurement, social construction, and enforcement of social categories. \looseness=-1
We then explore the constructs of fairness, robustness, and responsibility in the context of governance in and for responsible AI.
Together, these perspectives help us unpack how \emph{measurement acts as a hidden governance process} in sociotechnical systems. 
Understanding measurement as governance supports a richer understanding of the governance processes already happening in AI---responsible or otherwise---revealing paths to more effective interventions.\looseness=-1

\end{abstract}


\maketitle

\section{Measurement is everywhere (but usually implicit)}

In algorithmic\footnote{
	In this paper we flexibly use ``algorithmic systems'' or ``automated decision-making systems,'' more broadly  as computational systems, AI systems, or sociotechnical systems. All such terms are meant to capture the algorithmic or computational aspects, which may be based on various machine learning models, as well as to capture the broader context of use, which may include data pipelines, system design and implementation, and organizational use.\looseness=-1  
} 
systems, social phenomena are constantly being operationalized, 
from creditworthiness, gender, and race, to employee quality, community health, product quality, and attention, user content preferences,  language toxicity, relevance, image descriptions---themselves laden with cultural and social knowledge. 
In practice, the measurements of these phenomena are constrained by existing practices, data availability, and other problems of problem formulation \cite{passi2017data,passi2019problemformulation}. 
The process of creating measures of such phenomena is a version of what social scientists call \textit{measurement}.
But in algorithmic system development and data science, this {measurement process} is almost always implicit---from operationalizing the theoretical understandings of unobservable theoretical constructs (toxicity, quality, etc.) and connecting them
to observable data.
Here we build from the measurement framework put forward by Jacobs and Wallach \cite{jacobs2021measurement} to show 
where and how social, cultural, organizational, and political values 
are encoded in algorithmic systems, and furthermore how these encodings then shape the social world. 
Any attempt at meaningful ``responsible AI''
must consider our values (where and when they are encoded, implicitly or otherwise) and
fairness-related harms (where and why they emerge).
The lens of measurement makes such considerations possible, and moreover, reveals where governance already is playing a role in responsible AI---and where it could. \looseness=-1
\looseness=-1

\subsection{The measurement process reveals where fairness-related harms emerge}

Here we build directly on the measurement framework put forward by Jacobs and Wallach \cite{jacobs2021measurement}. 
Using this framework, we can understand concepts such as employee quality, language toxicity, community health, or social categories, as \textit{unobservable theoretical constructs} that are reflected in observable properties of the real world. To connect these observable properties to the constructs of interest, we \textit{operationalize} a theoretical understanding of the construct and, with a \textit{measurement model}, show how we connect observable properties to that operationalization. 
In practice, this process happens constantly in the design of sociotechnical systems. Sometimes this process of assumptions, operationalization and measurement are more explicit: {proxies} refer to a special case of measurement models. Sometimes the distinction between constructs, operationalizations, and measurements are more obscured: consider using inferred demographic attributes, predicting clickthrough rate to infer attention, or using other system data exhaust to predict behavior. 
\looseness=-1

This process matters beyond academic exercise and pedantry
because fairness-related harms emerge when there is a mismatch between the thing we purport to be measuring and the thing we actually measure
\cite{jacobs2021measurement,jacobs2020meaning}.
As an example, 
consider a tool for hiring or promotion: we might choose to operationalize `employee quality' with `past salary.' We could then choose the measurement model of `past salary' as an individual's average salary over the last three years. Even in this simple example, 
using this operationalization of employee quality would reinforce existing gender and racial pay disparities by systematically encoding `quality' as salary. This is an example of one such fairness-related harm---revealed by a mismatch between the construct of `employee quality' and its operationalization, `past salary.' 
\looseness=-1

\section{Measurement is governance}

We have thus far established that 
measurement is everywhere, whether we acknowledge it or not;
that fairness-related harms emerge from mismatches in the measurement process;
and that being explicit about about the measurement process, whether proactively or retrospectively, can reveal where those mismatches (and potential for harms) emerge.
A theme from this past work is that the \emph{assumptions} underlying sociotechnical systems are crucial, and that frameworks that reveal those assumptions can mitigate and prevent harms. 
Furthermore, assumptions are \emph{decisions}---and are decisions that reflect values---therefore frameworks that reveal those assumptions reveal implicit decisions in the creation, use, and evaluation of sociotechnical systems. 
So where is governance in responsible AI? \looseness=-1
Let us consider two sites. \looseness=-1

\paragraph{Governance in responsible AI}
First, decisions about what happens \emph{within} algorithmic systems are governance decisions. What data is used where, to what ends; what is being predicted or decided; what variables are named what and accessible to whom---all of these decisions shape what a system can and cannot do. Some of these decisions are explicit: for instance, gender is a binary variable; these advertisements should be served to those inferred income groups; only users that have selected one of those two binary gender variables will be eligible for an account; only users with a paid account can access this data; this functionality will be named `Friend' and this one, `Like'. The impacts of such decisions, of course, can extend well beyond their immediate implementation. And the downstream consequences of the decisions---of who is included, of who gets shown what content, of how we engage with a system through an API or as a user, of how we interpret functionalities (`Like')---need not have been explicitly engaged with.
This \emph{is} measurement---of gender; of relevance of content and of socio-economic status; of a complete or verified identity; of membership; of relationships and interactions. \looseness=-1

\paragraph{Governance for responsible AI}
The second site of governance is decisions \emph{about} algorithmic systems.
Is the system good at what it is supposed to do? Does it work the way we expect? (And specifically, does it perform well on the training set? Does it generalize to other settings?) Is it ready for deployment, and for which contexts? Or even more directly: Is our system fair? Is it robust? Is it responsible? 
Taking a step back, where does ``fair'' or ``robust'' or ``responsible'' come from? 
These are themselves essentially contested constructs that are being operationalized \cite{mittelstadt2019ai, mulligan2019thing, jacobs2021measurement}. Measurement is governance here too. \looseness=-1

\paragraph{Thinking tools} 
What of it then? This integrated perspective---measurement as governance---is a generative one.
We have tools---i.e., construct validity and reliability---to unpack the measurement process in algorithmic systems and uncover sources of potential fairness-related harms---i.e., mismatches in the measurement process \cite{jacobs2021measurement}. 
Two types of validity, content and consequential, are particularly well-suited to explore governance decisions: Content validity elicits substantive understanding of the construct being operationalized, \looseness=-1
while consequential validity reveals the feedback loop of governance decisions and their impacts as a part of validity. \looseness=-1
Other types of construct validity and reliability may already be familiar to data scientists and engineers alike, including face validity (do the assumptions seem reasonable?), test--re-test reliability (this includes out of sample testing: do we get similar results if we re-run the system?), and convergent validity (do the measures correlate with others of the same phenomenon?).  
These types may be already familiar under different names or within different frameworks of best practices. Thus an important caveat is that \emph{naming} which types of validity are being considered is not necessary to explore them. Asking the questions or interrogating assumptions, or assessing validity and reliability, can be done without the specific vocabulary. However, we can use the types of validity as inspiration for what questions we ought to ask ourselves and ask of our systems. 
\looseness=-1

\begin{figure}{} %
  \begin{center}
  \includegraphics[width=0.44\textwidth]{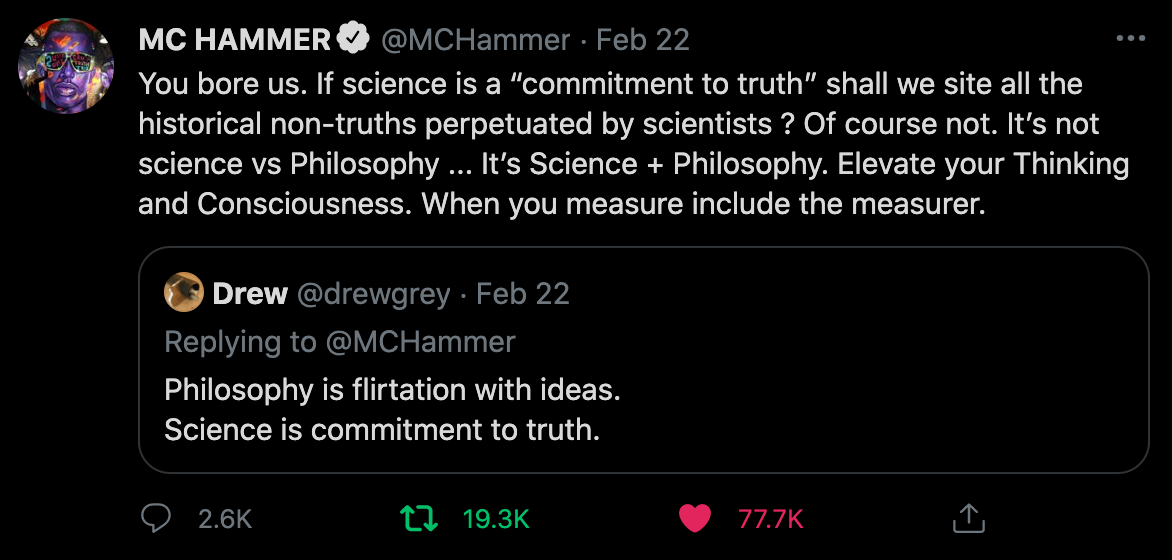} 
  \caption{
This broader systems-level view---of how measures are created and used---falls under consequential validity, that is, 
considering how measurements are used, and what assumptions were made by whom in the process.\looseness=-1
}
  \label{fig:measurer}
  \end{center}
\end{figure}

Finally, understanding measurement as governance 
serves as a reminder that these assumptions will be inherently shaped 
by those making them
 \citep[e.g.,][]{abebe2020roles, boyd2011six, gitelman2013raw}.
The contexts where these assumptions are made, and by whom, are not necessarily separable from the work itself	\cite{benjamin2019race,stevens2021seeing,passi2017data,passi2019problemformulation,goldenfein2018profiling,jafarinaimi2015values}. 
But we do better to articulate what assumptions are being made 
when, by whom, and to what downstream effects, when we consider this broader context 
(Figure \ref{fig:measurer}: ``when you measure, include the measurer'' \cite{hammertime}).\looseness=-1

\section{Validity and governance}

The tools that are available to interrogate assumptions (decisions) built in to sociotechnical systems, i.e., construct validity and reliability, can be used to interrogate governance practices. 
Furthermore, these tools can be used to connect with broader conversations on governance in and for AI. 
Here we explore several of those connections.
These settings---non-exhaustive and illustrative---point to the breadth of opportunities for rigorously considering the often-implicit governance decisions built into sociotechnical systems.
\looseness=-1

\subsection{Content and consequential validity}
Two types of validity are of particular relevance to governance in and for responsible AI: content validity and consequential validity.

\paragraph{Content validity}
{Content validity} captures two important places of mismatches:
first, is there a coherent, agreed upon theoretical understanding of the construct of interest? 
And second, given a theoretical understanding of that construct, does the operationalization fully match that understanding? 
Many constructs of interest will face the first challenge, and also vary across contexts, time, and culture
(fairness, for instance \citep[e.g.,][]{mulligan2019thing,hutchinson2019fiftyyears,sambasivan2021re,chouldechova2018frontiers}).
Even if there is not an agreed upon definition, \textit{something} is still being implemented: content validity then covers the substantive match between the operationalization, i.e., what is being measured, and the construct, i.e., what is intended to be measured. 
(Why bother with other types of validity then, if this apparently captures everything? Different types of validity, and the exploration one might pursue while trying to establish them, can reveal previously under- or mis-specified assumptions \cite{jacobs2021measurement}.) 
Content validity gets at both governance \textit{in} responsible AI---e.g., have we adequately captured ``creditworthiness'' or ``risk to society''?---and governance \textit{for} responsible AI---e.g., when can we declare our system to be ``fair'' or ``responsible''? 
This is a significant challenge---and even more so if that theoretical understanding is unstated or is understood differently by different stakeholders. 
For governance in AI, 
Passi and colleagues show how real-world organizations confront 
 both theoretical mismatches and mismatches between operationalization and construct throughout the multistakeholder process of making data science systems work \cite{passi2019problemformulation,passi2020making}. 
For governance for AI, debates about content validity are more explicit, both in defining responsible AI systems and in operationalizing legal guidelines for algorithmic systems
\citep[e.g.,][]{stark2021critical,greene2019better,mittelstadt2019ai,mittelstadt2016ethics,wachter2021fairness}.
\looseness=-1

\paragraph{Consequential validity}
Consequential validity
considers impact. This is not obviously a part of validity, in part because its name connotes a normative assessment. 
The argument that Samuel Messick, a researcher at the Educational Testing Service, put forward was that how measures are defined changes how they are used---therefore defining the measure changes the world it was created for \cite{messick1987validity,messick1996validity,jacobs2021measurement}. That is, consequential validity must be a part of validity, as a measure's operationalization changes how we understand that construct in the first place.
More recently, Hand writes that
``measurements both reflect structure in the natural world, and impose structure upon it'' \cite{hand2016measurement}. That is, it is clear that there is a \emph{feedback loop} between what we measure and how we interpret it: so to govern systems well, we must consider how the governance changes the system itself.
\looseness=-1

The feedback loop inherent in 
measurement is recognized from different perspectives
 across the social sciences. 
One notion is as Campbell's Law and its cousin 
Goodhart's Law \cite{campbell1979assessing,hand2016measurement}: ``when a measure becomes a target it ceases to be a good measure'' \cite{strathern1997improving}.
Messick, and the broader educational community, refer to the evocatively-named washback, where incentives around testing lead to teaching to the test \cite{messick1996validity}. 
In economics, these ideas appear through the Lucas critique and market devices \cite{hoover1994econometrics,lucas1976econometric,mackenzie2008engine,muniesa2007introduction},  where models of the economy and policies then shape the economy; similar ideas appear in sociology \cite{healy2015performativity}. \looseness=-1
Philosophical and sociological understandings of performativity explore how enacting categories changes their meaning \cite{hacking1999social} and how categories and value are socially co-constructed \citep[e.g.,][]{perry2004desi,ridgeway1991social,benjamin2019race,roberts2011fatal}.
This feedback loop has long been of interest in critical infrastructure studies and science and technology studies  \cite{star1999ethnography,bowker2000sorting,hacking1999social}.\footnote{From Haraway: ``Indeed, myth and tool mutually constitute each other.''~\cite{haraway2006cyborg}} 
\looseness=-1
Crucially, consequential validity formally brings systems-level feedback  
into our technical understanding of the measurement process. \looseness=-1

\subsection{Fairness, robustness, and responsibility}\label{subsec:substantive}

Rehashing the full breadth of 
disagreements about how constructs such as fairness are operationalized (in algorithmic systems specifically, and governance systems more broadly) is beyond the scope of this work. 
Here we point to some perspectives made available through the lens of construct validity before discussing governance in practice in section \ref{subsec:govern}.
\looseness=-1

\paragraph{Fairness}
Fairness is an essentially contested construct with diverse, conflicting, and context-dependent understandings of the theoretical construct \cite{mulligan2019thing,jacobs2021measurement,green2021impossibility}. 
Even with a given theoretical understanding of fairness, it is still nontrivial (and sometimes impossible) to 
operationalize the full theoretical scope of the intended definition 
\citep[e.g.,][]{hellman2020measuring,mitchell2021algorithmic,arneson2018four}, 
particularly  to implement ``this thing called fairness'' \cite{mulligan2019thing} 
 in meaningful applied settings (real-world organizations, legal systems) \cite{passi2019problemformulation,sambasivan2021re,holstein2018improving,wachter2021fairness}.
These are familiar challenges to the content validity of fairness: does the operationalization match the construct? \looseness=-1
However, content and consequential validity of fairness in algorithmic systems provide a wider view. 
In content and consequence, operationalizations of fairness almost always neglect notions of justice and power, and operationalizations based on static, narrow notions of fairness will perpetuate existing inequities \cite{kasy2021fairness,hoffmannrawls,green2018myth,hoffmann2019fairness}.
When encountering a ``fair'' system, however,
 different stakeholders may expect equitable or equal treatment, justice or dignity \cite{green2018myth,hoffmannrawls}. 
 The lens of consequential validity is key here: 
 A system that is labeled as ``fair'' under a narrow or static definition may further exacerbate inequities simply by labeling inequitable outcomes as ``fair.'' 
Breaking out of this feedback loop must then be done intentionally \cite{birhane2021algorithmic}.
 \looseness=-1

\paragraph{Robustness}
Robustness---asking how well a system performs, perhaps across contexts, or to do the thing it claims---may seem more neutral.
Is the system good at what it is supposed to do? Does it work the way we expect? Is it ready for deployment? (And specifically, does it perform well on the training set? Does it generalize?) 
A growing literature has turned to 
the choices involved in how organizations evaluate these assessments of robustness
\cite{passi2020making} and, more broadly,
the politics of seemingly technical decisions about model performance \cite{hancox2021epistemic,chasalow2021representativeness}, 
into 
the politics of developing larger data sets and models in the name of robustness 
\cite{birhane2021large,paullada2020data,bender2021dangers,stevens2021seeing,denton2020bringing}.
Through the lens of consequential validity (i.e., it matters what we call things), implying that robustness only requires technical decisions obscures the hidden governance decisions at hand \cite{birhane2021algorithmic,chasalow2021representativeness,hancox2021epistemic}. \looseness=-1

\paragraph{Responsibility}
Finally, we can turn to responsibility. 
Without adjudicating different desirable properties of responsible algorithmic systems, 
we point to several threads:
first, that responsibility is about harms. Centering harms foregrounds risk and power. Critical scholars have shown how pivoting 
 to a human rights perspective, away from ethics and fairness, allows us to fundamentally attend to (mitigating) fairness-related harms 
\cite{cath2020leap,birhane2020robot,greene2019better,birhane2021algorithmic}. 
Second, naive attempts towards inclusion create opportunities for further harms
 \cite{hoffmann2020terms,keyes2018misgendering,bennett2020point}. As with robustness, moving towards responsible systems through better auditing  
benchmarks provides an awkward path toward accountability \cite{denton2020bringing,stevens2021seeing,raji2020saving}. 
We now turn to ways that so-called responsible AI systems are governed more broadly, drawing from perspectives from measurement and construct validity.
\looseness=-1

\subsection{Governing (responsible) AI}\label{subsec:govern}

The development and critique of governance in and for AI already uses the many of the perspectives discussed here. 
This is in no small part due to the framing of the challenge. If we begin from a focus on algorithmic systems in which measurement processes are mostly implicit, we can then use the perspectives given by measurement and construct validity to uncover the hidden governance decisions in algorithmic systems.
However, 
if we begin from a focus on governance, 
then the relevance and use of the
perspectives given by measurement and construct validity are self-evident: what do we want our system to achieve? What values or principles do we want to enact? How do we know if our system is upholding those values? What potential harms may emerge if we fail? (Of course, answering these questions remains a challenge.)
Furthermore, the potential for harms arising from mismatches between the constructs intended to be operationalized and their operationalization is also more prominent: in 2016 Mittelstadt and colleagues wrote that the
``gaps between the design and operation of algorithms and our understanding of their ethical implications can have severe consequences affecting individuals as well as groups and whole societies'' \cite{mittelstadt2016ethics}.
As such, 
efforts to operationalize responsible governance for responsible AI
have focused on developing values, principles, and ethics statements (both widely discussed elsewhere 
\cite{cath2018governing, floridi2018soft, floridi2019establishing} and thoughtfully critiqued \cite{mittelstadt2019ai,birhane2021algorithmic, stark2021critical}).
While measurement perspectives may be more readily available, 
the challenges to meaningfully operationalize responsible AI are nontrivial in practice
\citep[e.g.,][]{rakova2020responsible,raji2020closing,madaio2020co,holstein2018improving,fish2020reflexive}.
 With deep ambiguity involved in converting values into practice 
\cite{jafarinaimi2015values,mittelstadt2019ai}, governance for AI needs the social sciences, interdisciplinary and diverse teams, and broader systems perspectives  \cite{sloane2019ai,mulligan2019thing,birhane2021algorithmic,green2020algorithmic}.

An emphasis on measurement can unify existing work on governance.
For instance, governance in the public sector relies on implicit measurement decisions that then shape policy
\cite{levy2021algorithms}, and  
governance tools like impact assessments operationalize harms and what must be done to mitigate them, i.e., impacts and assessments thereof are co-constructed 
\cite{metcalf2021algorithmic}.
The consequences of governance decisions play out in other domains as administrative or symbolic violence---for instance
when technical systems classify individuals in a way that misrepresent and exclude trans and nonbinary individuals \cite{spade2015normal,keyes2018misgendering} or individuals with disabilities \cite{hutchinson2020unintended,bennett2020point}.
Consequential validity specifically helps account for how meaning is made, encoded, and enforced
 (Alder: ``measures are more than a creation of society, they create society''~\cite{alder2002measure}).
We can then functionally connect our understanding of consequential validity to large and nearby conversations: for instance, race as and of technology \cite{benjamin2019race,roberts2011fatal}, from the larger context of critical race theory \cite{delgado2017critical,hanna2020towards}. 
Similarly, by shifting the focus to {harms}---rather than, for instance, bias---we have the opportunity to reflect on what critical assumptions are being made; to reflect on what impacts are being addressed (unequal outcomes may be more or less equitable); to reflect on \emph{power} (including from where decisions are being enforced, by whom, and who they impact); 
and to reflect on what social phenomena are being displayed---and for which the fixes may not be technical
\cite{jacobs2020meaning,blodgett2020language,birhane2021algorithmic,sloane2020participation}.\looseness=-1
Finally, explicitly connecting measurement and governance
makes relevant lessons from past governance successes (and failures) 
in algorithmic systems as well as other complex social and organizational systems. \looseness=-1

\section{Conclusion}

Developing ``responsible AI'' is and will be a social, technical, organizational, and political activity.
Angela Carter in \emph{Notes from the Front Line} writes that ``Language is power, life and the instrument of culture, the instrument of domination and liberation.''
In algorithmic systems, it is the often-implicit measurement process where this instrumenting---where governance---happens. 
By understanding measurement as governance, we can bring  critical and social science 
perspectives more formally into our conceptions of responsible AI \cite{stark2021critical,sloane2019ai,mulligan2019thing,passi2017data}, while equipping researchers and practitioners with better tools and perspectives to develop responsible AI.
\looseness=-1


\end{document}